\documentclass[
 aip, 
 amsmath,amssymb,
 reprint,%
]{revtex4-1}

\usepackage{graphicx}
\usepackage{dcolumn}
\usepackage{bm}

\usepackage[utf8]{inputenc}
\usepackage[T1]{fontenc}
\usepackage{mathptmx}
\DeclareMathAlphabet{\altmathcal}{OMS}{cmsy}{m}{n}
\usepackage{cancel}
\begin{document}

\preprint{AIP/123-QED}

\title[Potential Vorticity Transport in Weakly and Strongly 
Magnetized Plasmas]{Potential Vorticity Transport in Weakly and Strongly 
Magnetized Plasmas}

\author{Chang-Chun Chen}
 \email{chc422@ucsd.edu }
\author{Patrick H. Diamond}%
 \email{pdiamond@ucsd.edu }
\author{Rameswar Singh}
\affiliation{ 
Department of Physics, University of California San Diego
}%

\author{Steven M. Tobias}
\affiliation{%
Department of Applied Mathematics, University of Leeds
}%

\date{\today}

\begin{abstract}
Tangled magnetic fields, often coexisting with an ordered mean field, have a major impact on turbulence and momentum transport in many plasmas, including those found in the solar tachocline and magnetic confinement devices.
We present a novel mean field theory of potential vorticity mixing in $\beta$-plane magnetohydrodynamic (MHD) and drift wave turbulence.
Our results show that mean-square stochastic fields strongly reduce Reynolds stress coherence.
This decoherence of potential vorticity flux due to stochastic field scattering leads to suppression of momentum transport and zonal flow formation.
A simple calculation suggests that the breaking of the shear-eddy tilting feedback loop by stochastic fields is the key underlying physics mechanism.
A dimensionless parameter that quantifies the increment in power threshold is identified and used to assess the impact of stochastic field on the L-H transition.
We discuss a model of stochastic fields as a resisto-elastic network. 
\end{abstract}

\maketitle
\begin{quotation}

\end{quotation}
\section{Introduction}
Momentum transport and the formation of sheared flows (i.e. zonal jets) are major research foci in quasi two-dimensional (2D) fluids\citep{pedlosky1979,bracco1998spotted} and plasmas\citep{mcintyre2003,diamondetal2005,Keating_Diamond2007, Chen_2020}. 
By `quasi 2D', we mean systems with low effective Rossby number, in which dynamics in the third dimension is constrained by, say, stratification or fast time averaging, due to small electron inertia (as in magnetically confined plasmas). 
In such systems, Reynolds forces are equivalent to vorticity fluxes via the Taylor Identity\citep{taylor1915}.
For this and other reasons---the most fundamental being the freezing-in law for fluid vorticity\citep{poincare1893}---it is natural to describe such systems in terms of \textit{potential vorticity} (PV). 
More generally, $PV \equiv \zeta = \underline{\zeta}_a \cdot \underline{\nabla} \psi/\rho$, where $\zeta_a$ is the absolute vorticity, $\psi$ is a conserved scalar field , and $\rho$ is the fluid density.    
The advantage of a PV description of the dynamics is that $\zeta$ is conserved along fluid particle trajectories, up to viscous dissipation, much likes phase space density is conserved in the Vlasov plasma. 
Examples of conserved PV are $\zeta =\beta y - \nabla^2 \psi$, where $\beta $ is Rossy parameter and $\psi$ is stream function, for dynamics on $\beta$-plane, and $PV =(1-\rho_s^2 \nabla^2)|e|\phi /T + \ln{n_0}$ for the Hasegawa-Mima system\citep{hasegawa1978pseudo}, where $\phi$ is electric potential and $n_0$ is a background density. 
In such systems, momentum transport and flow formation are determined by inhomogeneous PV mixing\citep{leprovost2007effect,wood2010general}. 
The mechanism for PV mixing is closely related to the coherence and cross phase of the vorticity flux. 
Mechanisms include viscous dissipation, wave-flow resonance, nonlinear mode interaction, and beat wave-flow interaction, akin to nonlinear Landau damping\citep{landau1946vibrations}.

Recently the physics of PV transport in a disordered magnetic field has emerged as a topic of interest in many contexts.
One of these is the solar tachocline\citep{Chen_2020}, a weakly magnetized system, where momentum transport (i.e. turbulent viscosity) is a candidate mechanism for determining the penetration of this layer and the flows within it.
The latter is critically important to the solar dynamo\citep{Parker:1993,Gruzinov1996,diamondetal2005}. 
In this case, the field is disordered, and confined (hydrostatically) to a thin layer.
The disorder magnetic field is amplified by high magnetic Reynolds number ($Rm$) turbulent motions\citep{Gruzinov1996,diamondetal2005}, pumped by convective overshoot from the convective zone\citep{fyfe_montgomery_1976, Brummell_2008}. 
There is a weak mean toroidal field, so magnetic perturbations are large.
Another application, relevant to PV dynamics in a stochastic magnetic field, is to tokamaks (strongly magnetized), specifically those with stochasticity induced by Resonant magnetic perturbations (RMPs)\citep{Evans_2015}.
RMPs are applied to the edge of tokamak plasma to mitigate Edge Localized Modes (ELMs)\citep{Evans_2005,Evans_2008}, which produce unacceptably high transient heat loads on plasma-facing components. 
The `cost' of this benefit is an increase in the Low to High confinement mode transition (L-H transition) threshold power, as observed with RMPs\citep{Leonard_1991,Gohil_2011,Kaye_2011, Ryter_2013, Mordijck_2015,Scannell_2015, in2017,Schmitz_2019}. 
Because several studies suggest that the L-H transition is triggered by edge shear flows\citep{diamond1994self, kim2003mean, malkov2009weak, estrada2011}, it implies that the transition dynamics are modified by the effects of stochastic fields on shear flow evolution.
Indeed, analysis suggests that RMPs may ``randomize" the edge layer. 
In this case, the magnetic field is three dimensional (3D).
Stochasticity results from $\underline{k} \cdot \underline{B} =0$ resonance overlap, and field line separations diverge exponentially. 
Hence, a key question is the effect of stochastic fields on self-generated shear flows. 


In both cases, the central question is one of phase---i.e. the effect of the stochastic field on the coherence of fluctuating velocities, which enters the Reynolds stress and PV. 
In physical terms, the disordered field tends to couple energy from fluid motion to Alfv\'enic and acoustic waves, which radiate energy away and disperse wave packets. 
Of course, Alfv\'enic radiation is more effective in the case for low $\beta \equiv p_{\text{plasma}}/p_{\text{mag}}$---the ratio of the plasma pressure to the magnetic pressure---or for incompressible dynamics. 
The effect of this Alfv\'enic coupling is to induce the decoherence of the Reynolds stress (or vorticity flux), thus reducing momentum transport and flow generation.  
In this vein, we show that sufficiently strong coupling of drift waves to a stochastic magnetic field can break the `shear-eddy tilting feedback loop', which underpins flow generation by modulational instability. 
We note that the interaction of Alfv\'en waves with a tangled magnetic field differs from that of Alfv\'en waves with an ordered field. 
Here, the effect is to strongly couple the flow perturbations to an effective elastic medium threaded by the chaotic field. 

In this paper, we discuss the theory of PV mixing and zonal flow generation in a disordered magnetic field, with special focus on applications to momentum transport in the solar tachocline and Reynolds stress decoherence in the presence of a RMP-induced stochastic field.
Section \ref{sec1} addresses a mean field theory for a tangled `in-plane' field in $\beta$-plane magnetohydrodynamic (MHD)\citep{Moffatt1978, gilman2000}, which is used to compute the Reynolds force and magnetic drag in this weak mean field ($B_0$) system.
The mean-square stochastic magnetic field ($\overline{B_{st}^2}$) was shown to be the dominant element, controlling the coherence in the PV flux and Reynolds force\citep{Chen_2020}. 
Of particular interest is the finding that the Reynolds stress degrades for weak $B_0$, which is well below that required for Alfv\'enization.
It is also shown that the small-scale field defines an effective Young's modulus for elastic waves, rather than a turbulent dissipation\citep{Chen_2020}. 
As a second application, Section \ref{sec:DWT} presents the study of Reynolds stress decoherence in tokamak edge turbulence. 
There, the stochastic field is 3D, and induced by external RMP.
Drift-Alfv\'en wave propagation along stochastic fields induces an ensemble averaged frequency shift that breaks the `shear-eddy tilting feedback loop'.
Reynolds stress decoherence occurs for modest level of stochasticity.
The ratio of the stochastic broadening effect to the natural linewidth defines a critical parameter that determines the L-H transition power threshold concomitant increment. 
With intrinsic toroidal rotation in mind, we also explore the decoherence of the parallel Reynolds stress.
This is demonstrated to be weaker, since the signal propagation speed which enters parallel flow dynamics is acoustic (not Alfv\'enic). 
The interplay of symmetry breaking, stochasticity, and residual stress are discussed. 
In Section \ref{sec: discussion}, we discuss the key finding of this study and provide suggestions for further research. 

\section{$\beta$-plane MHD and the Solar Tachocline}\label{sec1}
\begin{figure}[h!]
\centering
  \includegraphics[scale=0.13]{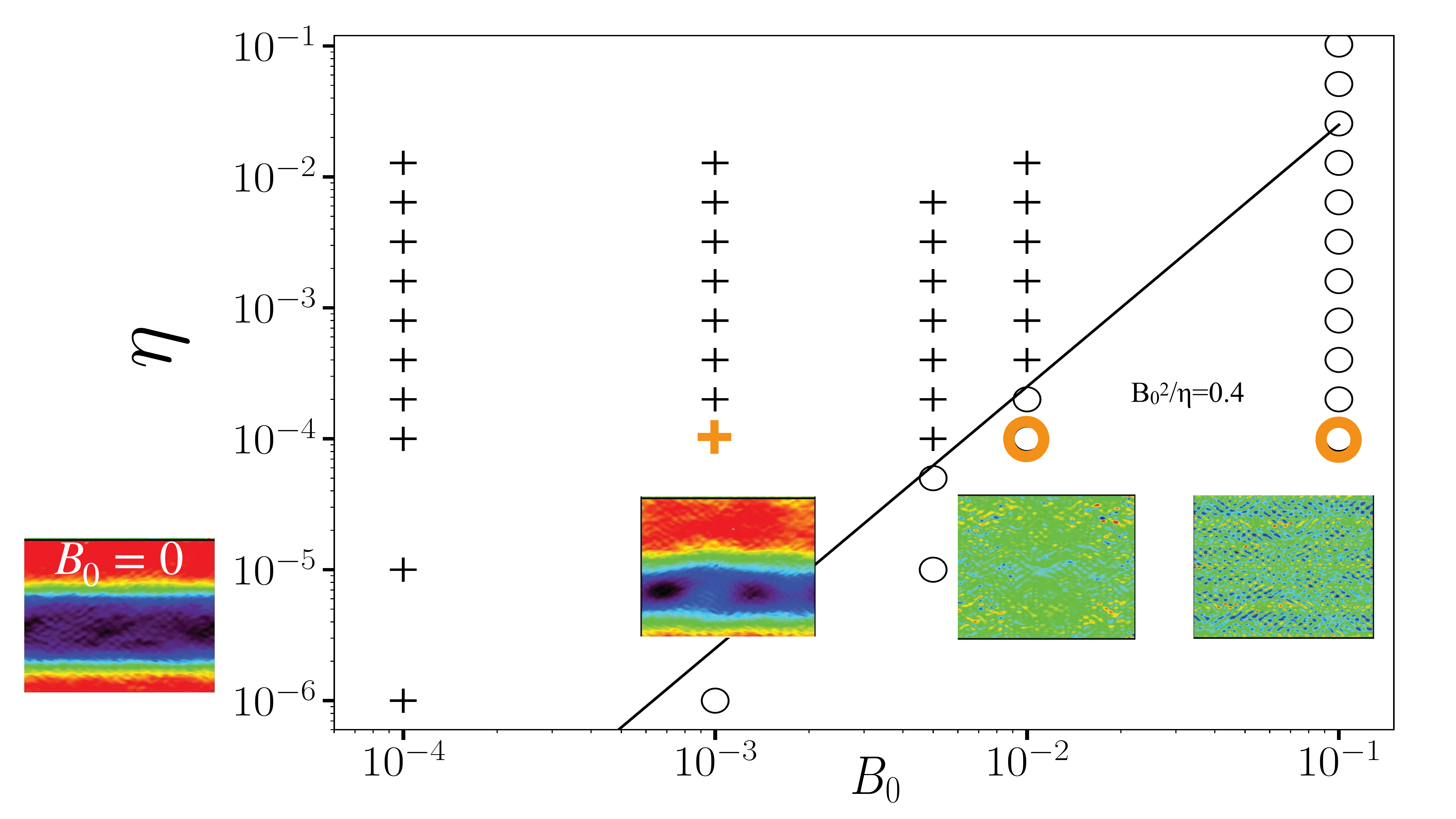}
  \caption{Scaling law for the transition between the forward cascades (circles)
and inverse cascades (plus signs) from \citet{tdh2007}. $B_0$ is mean magnetic field.
	Colormaps are velocity intensity. Red indicates strong forward flows, while blue indicates strong backward flows.
	They shows as mean magnetic field strong enough, zonal flow is ceased and the system is fully Alfv\'enized. 
	}
  \label{fig:etal}
\end{figure}

Stochastic fields are ubiquitous.
One example is the tangled field of the solar tachocline\citep{cdthom2007,Chen_2020}---a candidate site for the solar dynamo. 
The solar tachocline is a thin strongly stratified layer between the radiation and convection zones, located at $\sim 0.7 $ solar radius\citep{cdthom2007}, where magnetic fields are perturbed by `pumping' from the convection zone.
Hence, a model for strong perturbed magnetic fields is crucial for studying PV mixing and momentum transport in the solar tachocline. 
A study by \citet{tdh2007} on $\beta$-plane MHD shows that a modest mean field suppresses zonal flow formation and momentum transport (Fig. \ref{fig:etal}).
\citet{Chen_2020} proposed that the effects of suppression by random-fields are already substantial (even for weak $B_0$) on account of Reynolds stress decoherence.
They discussed a $\beta$-plane (quasi-2D) MHD model for the solar tachocline and studied how the zonal flow is suppressed by random fields.
We note that the dynamics of $\beta$-plane MHD are exceedingly complex. 
At small-scales, it resembles MHD with a forward cascade and also supports large scale Rossby waves.
Interactions of the latter tend to generate flows, as for an inverse cascade.
In view of this multi-scale complexity, we follow the suggestion of \citet{Rosenbluth1977} and replace the full problem by a more treatable one in which an ambient disordered field is specified. 
We utilize a mean field theory which averages over the small-scale field. 
Meso-scopic flow phenomena in this environment are then examined. 
\subsection{Model Setup}
\begin{figure}[h!]
\centering
  \includegraphics[scale=0.25]{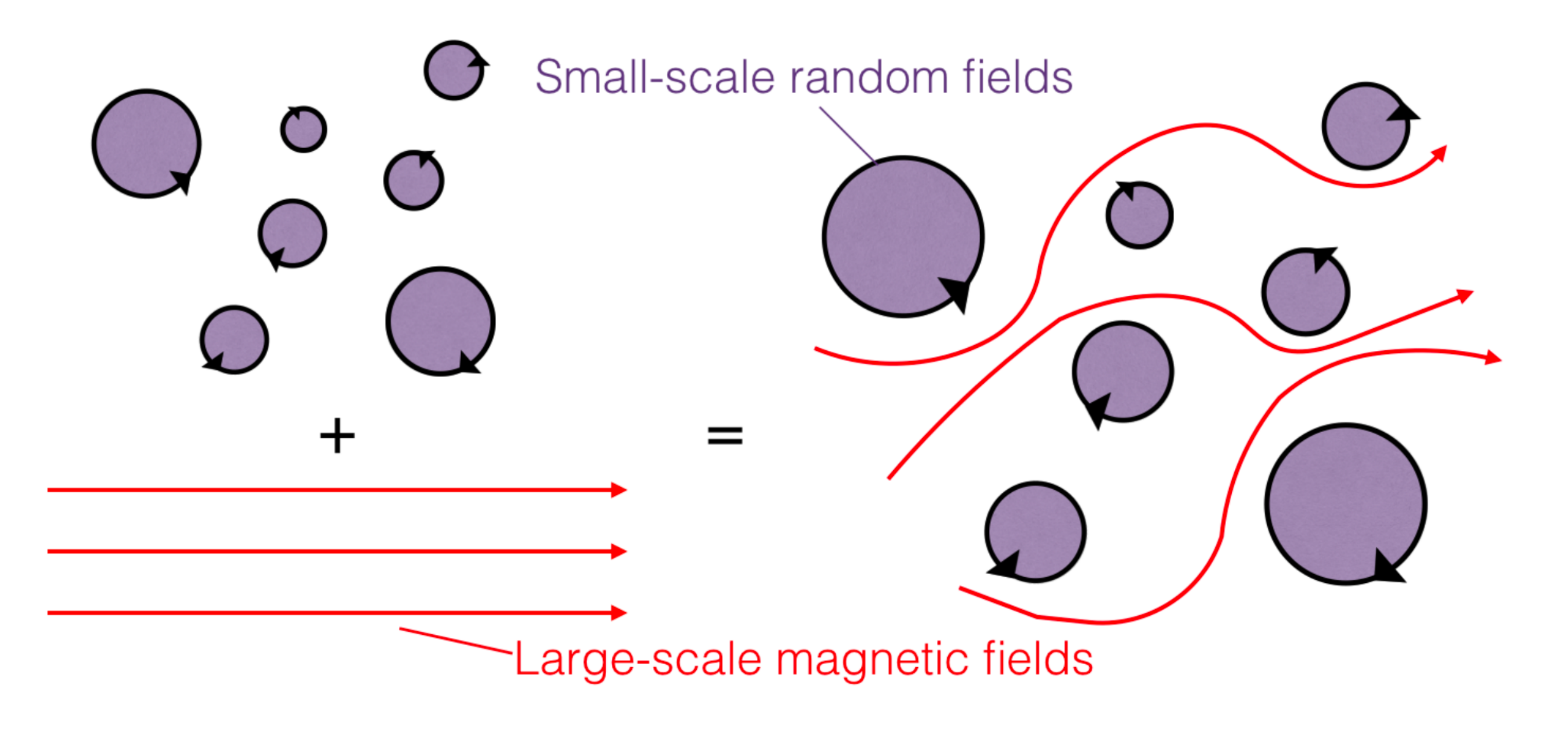}
  \caption{The large-scale magnetic field is distorted by the small-scale fields.
  The system is the `soup' of cells threaded by sinews of open field lines. }
  \label{fig:channel}
\end{figure}

The $\beta$-plane MHD system at high $Rm$ with \textit{weak} mean field supports a strong disordered magnetic field.
Hence, analyzing this problem is a daunting task, on account of the chaotic field and strong non-linearity.  
\citet{Zel1983_mag_percolation} suggested the `whole' problem consists of a random mix of two components: a weak, constant field ($B_0$) and a random ensemble of magnetic `cells' ($B_{st}$), for which the lines are closed loops ($\nabla \cdot \bold{B_{st}} = 0$).
Assembling these two parts gives a field configuration which may be thought of as randomly distributed `cells' of various sizes, 
threaded by `sinews' of open lines (Fig. \ref{fig:channel}). 
Hence, the magnetic fields can be decomposed to $\bold{B} \equiv \bold{B_0} + \bold{B_{st}}$, where $ B_0$ is modest (i.e. $|B_{st}| > B_0$). 
This system with strong, tangled field cannot be described by linear responses involving $B_0$ only, and so is not amenable to traditional quasilinear theory. 
Linear closure theory allows analysis in a diffusive regime, where fluid Kubo number\citep{kubo1963} $Ku_{fluid} <1$ and magnetic Kubo number $Ku_{mag} < 1$. 
For weak mean field, we have $Ku_{mag}\equiv l_{ac} |B_{st}/B_0|/\Delta >1$, rendering standard closure method inapplicable. 
Here $l_{ac} $ is magnetic auto-correlation length and $\Delta$ is eddy size.
Hence, we employ the simplifying assumption of $l_{ac}\rightarrow 0 $ so $Ku_{mag}\simeq l_{ac} |B_{st}/B_0|/\Delta <1$.
This approximation allows us to peek at the mysteries of the strong perturbation regime by assuming delta-correlated fields. 
In a system with strong random fields ($B_{st}$; such that ensemble average of squared stochastic magnetic field $\overline{B_{st}^2} > B_0^2$), this approximation comes at the price of replacing the full $\beta$-plane MHD problem with a model problem. 
Results for this model problem, where $|B_{st}| > B_0$, are discussed.  
\subsection{Calculations and Results} \label{sec: Cal. Re. of beta-plane}
\begin{figure}[h!]
\centering
  \includegraphics[scale=0.15]{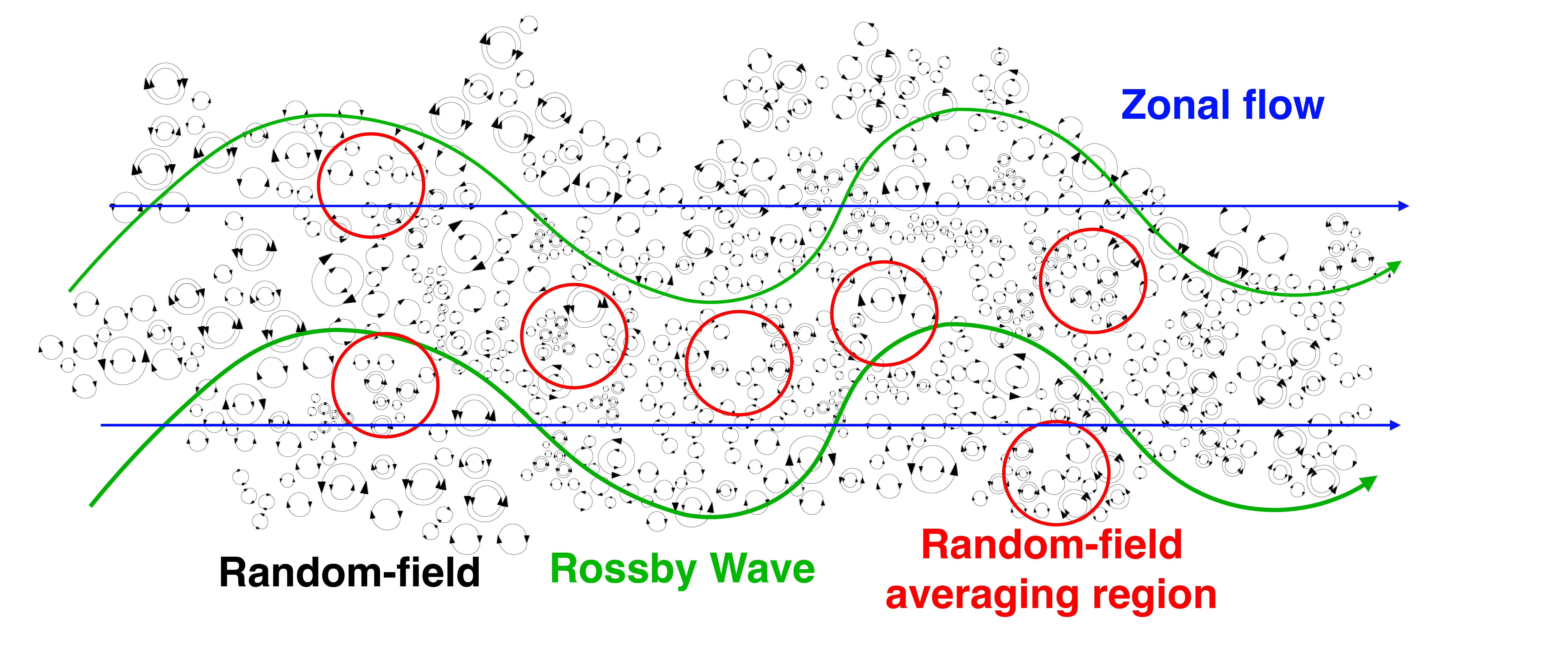}
  \caption{Length scale ordering. The smallest length scale is that of the random field ($l_{st}$). The random-field averaging region is larger than the length scale of random fields but smaller than that of the Rossby waves. }
  \label{fig:averag_scale}
\end{figure}

Following the argument above, a model which circumvent the problem of simple quasi-linear theory for this highly disordered system is presented.
This is accomplished by considering the scale ordering.
In the \textit{two-scale average method} proposed\citep{Chen_2020}, an average over an area is performed, with a scale ($1/k_{avg}$) larger than the scale of the stochastic fields ($1/k_{st}$) but smaller than the Magnetic Rhines scale \citep{rhines_1975} ($k_{MR}$), and Rossby wavelength ($k_{Rossby}$).
This average is denoted as 
$
\overline{F} \equiv \int dR^2 \int dB_{st} \cdot P_{(B_{st,x},B_{st,y})} \cdot F
$,
where $F$ is arbitrary function, $dR^2 $ denotes integration over the region, and $P_{(B_{st,x},B_{st,y})}$ is probability distribution function for the random fields. 
This random-field average allows us to replace the total field due to MHD turbulence (something difficult to calculate) by moments of a prescribed probability distribution function (PDF) of the stochastic magnetic field.
The latter \textit{can} be calculated. 
Another ensemble average--- over zonal flow scales $k_{zonal}$, denoted as bracket average $\langle \rangle \equiv \frac{1}{L} \int dx \frac{1}{T}\int dt$---is conducted. 
Hence the scale ordering is ultimately $k_{st}>k_{avg} \gtrsim k_{MR} \gtrsim  k_{Rossby}>k_{zonal}$ (Fig. \ref{fig:averag_scale}).
This model\citep{Chen_2020} with its two-average method allows insights into the physics of how the evolution of zonal flows is suppressed by disordered fields both via reduced PV flux ($\Gamma$) and by an induced magnetic drag, i.e. 
\begin{equation}
	\frac{\partial}{\partial t}  \langle u_x \rangle
	=  \langle
	 \overline{\Gamma} 
	 \rangle
	- \frac{1}{ \eta\mu_0\rho}  
	\langle \overline{B_{st,y}^2} \rangle  \langle u_x\rangle + \nu \nabla^2 \langle u_x 
	\rangle.
	\label{zonal_sup_PV_drag}
\end{equation}
Here, $\langle u_x \rangle$ is mean velocity in the zonal direction, $\langle\overline{\Gamma}  \rangle$ is the double-average PV flux, $\eta$ is resistivity, $\rho$ is mass density, and $\nu$ is viscosity.
Here $\frac{1}{ \eta\mu_0\rho}  
	\langle \overline{B_{st,y}^2} \rangle   $ is the magnetic drag coefficient.
	
\begin{figure}[h!]
\centering
  \includegraphics[scale=0.13]{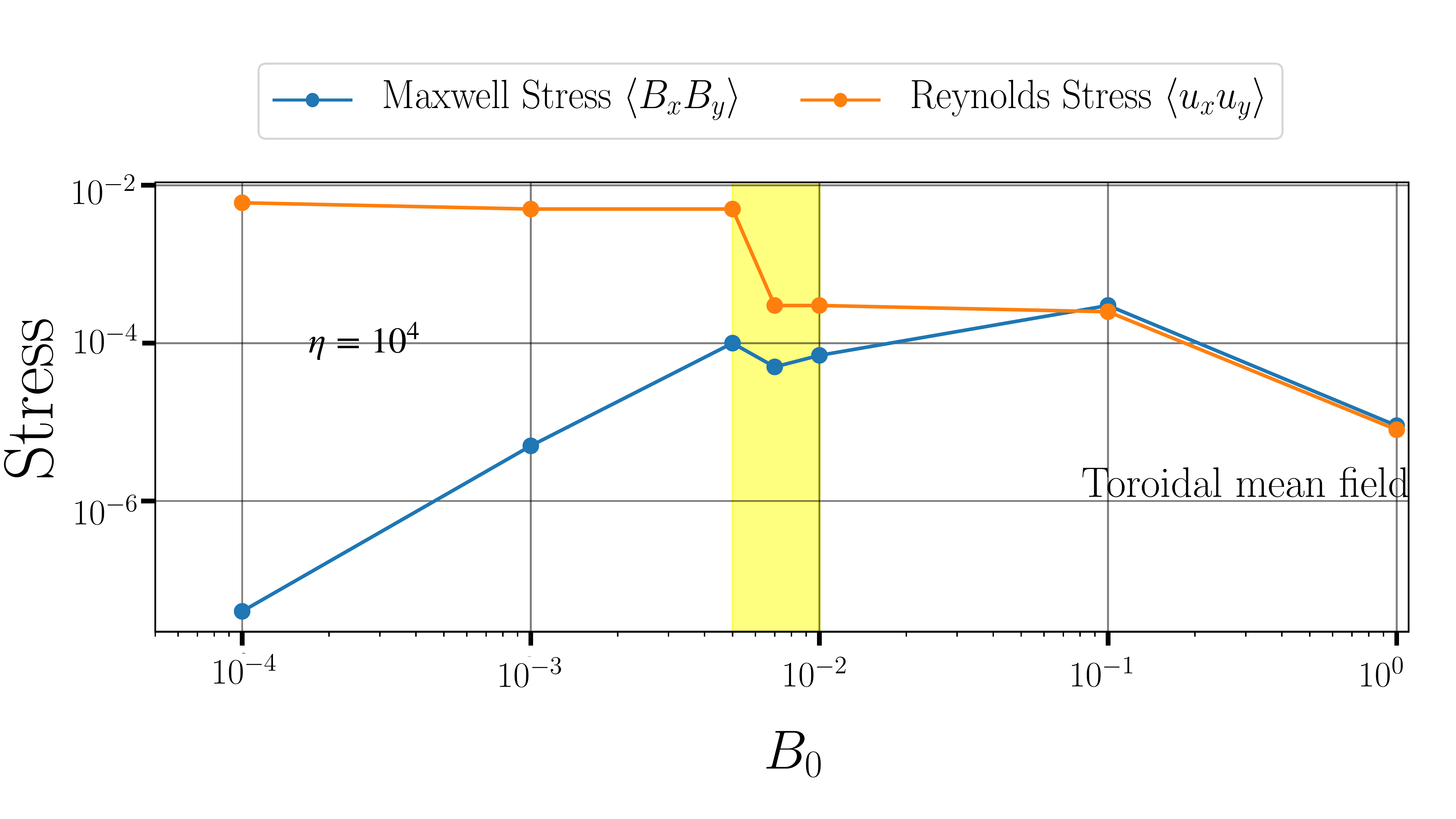}
  \caption{ Average Reynolds stresses (orange line) and Maxwell stresses (blue line) for $\beta$ = 5, $\eta = 10^{-4} $ from \citet{Chen_2020}.
	Full Alfv\'enization happens when $B_0$ intensity is larger than $B_0 = 10^{-1}$ and $B_0 = 6 \times 10^{-2}$, respectively. 
	The yellow-shaded area is where zonal flows cease to grow.
	This is where the random-field suppression on the growth of zonal flow becomes noticeable. }
  \label{fig:stresses}
\end{figure}

First, stochastic fields suppress PV flux by reducing the PV diffusivity ($D_{PV}$)
\begin{equation}
	\overline{\Gamma} = - D_{PV}		\big(\frac{\partial}{\partial y} \overline{\zeta} + \beta \big),
\end{equation}
where $\beta$ is the Rossby parameter and the PV diffusivity can be written as 
\begin{equation}
\begin{split}
	&D_{PV} = \sum_{k} |\widetilde{u}_{y,k} |^2 \times
\\
	&\frac{\nu k^2 
		+(\frac{B_0^2 k_x^2}{\mu_0 \rho})\frac{ \eta k^2} {\omega^2 + \eta^2 k^4} 
		+  \frac{ \overline{B_{st,y}^2} k ^2 }{\mu_0 \rho  \eta k^2} 
		}
		{\bigg( \omega- (\frac{B_0^2 k_x^2}{\mu_0 \rho})\frac{\omega}{\omega^2 + \eta^2 k^4} \bigg)^2	
		+\bigg(
		\nu k^2 
		+ (\frac{B_0^2 k_x^2}{\mu_0 \rho})\frac{ \eta k^2}{\omega^2  + \eta^2 k^4} 
		+ \frac{ \overline{B_{st,y}^2} k ^2 }{\mu_0 \rho  \eta k^2}
		\bigg)^2.
		}
\end{split}
\label{eq: eq3}
\end{equation}
Eq. (\ref{eq: eq3}) shows that strong mean-square stochastic field ($\overline{B_{st}^2}$) acts to reduce the correlation of the vorticity flux, thus reducing PV mixing. 
 This explains the Reynolds stress suppression observed in simulation\citep{Chen_2020} (Fig. \ref{fig:stresses}).
Note that this reduction in Reynolds stress sets in for values of $B_0$ \textit{well below that required for Alfv\'enization} (i.e. Alfv\'enic equi-partition $\langle \widetilde{u}^2\rangle  \simeq \langle \widetilde{B}^2 \rangle/\mu_0 \rho$).

\begin{figure}[h!]
\centering
  \includegraphics[scale=0.17]{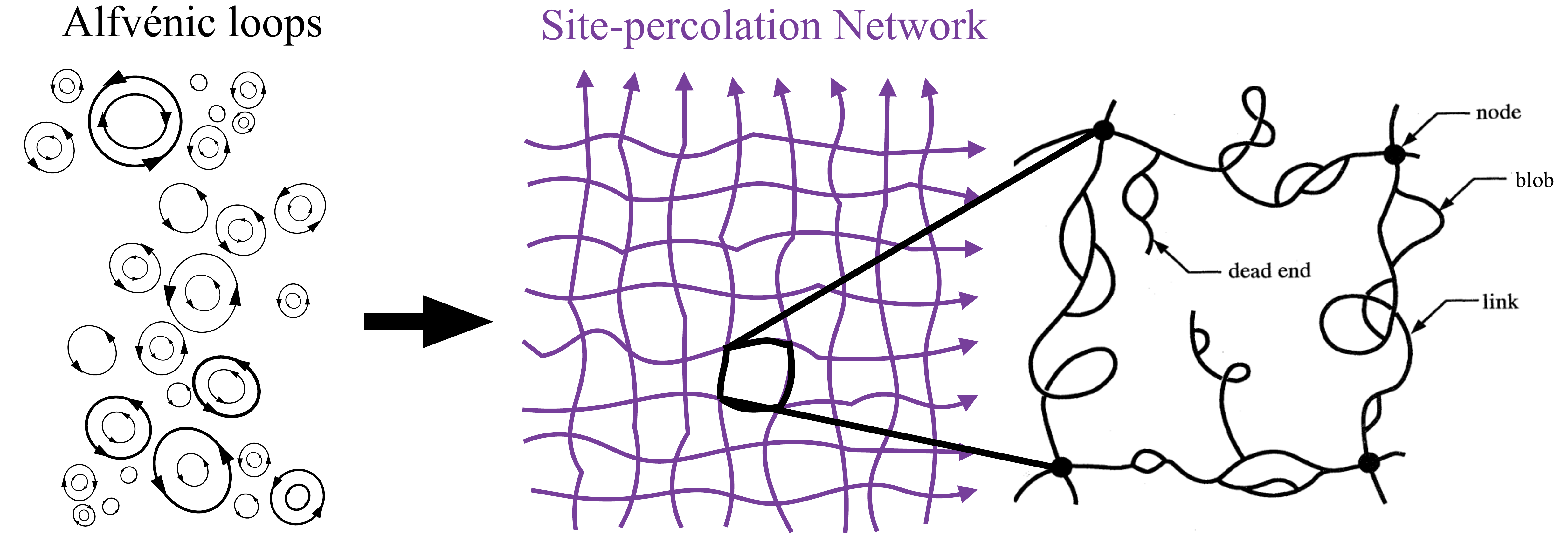}
  \caption{Site-Percolation Network.
	Schematic of the nodes-links-blobs model (or SSdG model, see \citealt{ Skal74, de1976relation, naka1994}).
	This depicts the resisto-elastic medium formed by small-scale stochastic fields.  }
  \label{fig:alfvenic_loop}
\end{figure}

Second, magnetic drag physics is elucidated via the mean-field dispersion relation for waves in an inertial frame ($\beta=0$), on scales $l \gg k_{avg}^{-1}$,
\begin{equation}
	 \bigg(  
	 \omega 
	 + \frac{i \overline{B_{st,\, y}^2} k_y^2 }{\mu_0 \rho \eta k^2}
	 + i\nu k^2 
	 \bigg) 
	 \bigg(\omega +  i \eta k^2  \bigg) 
	 = 
	 \frac{B_0^2 k_x^2 }{\mu_0 \rho}.
\end{equation}
The drag coefficient $\chi \equiv \frac{ \overline{B_{st,\, y}^2} k_y^2 }{\mu_0 \rho \eta k^2}$, emerges as approximately proportional to an effective $\frac{\text{spring  constant}}{\text{dissipation}}$.
The `dissipation' and `drag' effects suggest that mean-square stochastic fields $\overline{B_{st}^2}$ form an effective resisto-elastic network, in which the dynamics evolve.
The fluid velocity is redistributed by the drag of small-scale stochastic fields.
Ignoring viscosity ($\nu \rightarrow 0$), we have
\begin{equation}
\omega^2 
	+ i\underbrace{(\chi +\eta k^2)}_{\text{drag + dissipation}}\omega 
	-\underbrace{\left(\frac{ \overline{B_{st, \, y}^2} k_y^2}{ \mu_0\rho} + \frac{ B_0^2 k_x^2}{ \mu_0\rho }  \right) }_{\text{effective spring constant}}
	=0.
\end{equation}
Note that this is effectively the dispersion relation of dissipative Alfv\'en waves, where the `stiffness' (or magnetic tension) is determined by \textit{both} the ordered and the mean-square stochastic field ($\overline{B_{st}^2}$).
In practice, the latter is dominant, as $\overline{B_{st}^2} \simeq Rm B_0^2$ and $Rm \gg 1$. 
So, the ensemble of Alfv\'enic loops can be viewed as an network of springs (Fig. \ref{fig:alfvenic_loop}).
Fluid couples to network elastic elements, thus exciting collective elastic modes.
The strong elasticity, due to Alfv\'enic loops, increases the effective memory of the system, thus reducing mixing and transport and ultimately causes Reynolds stress decoherence. 
The network is fractal and is characterized by a  `packing factor', which determines the effective Young's Modulus. 
It is important to note that the `stochastic elasticized' effect is one of increased memory (\textit{not} one of enhanced dissipation) as in the familiar cases of turbulent viscosity or resistivity. 

\subsection{Implications for the solar tachocline}
The balance between Reynolds and Maxwell stress in a fully Alfv\'enized system where fluid and magnetic energy reach near equi-partition is the conventional wisdom.  
Simulation results (Fig. \ref{fig:stresses}), however, show that Reynolds stress is suppressed by stochastic fields \textit{well before} the mean field is strong enough to fully Alfv\'enize the system \citep{Chen_2020}.
These results suggest that turbulent momentum transport in the tachocline is suppressed by the enhanced memory of stochastically induced elasticity. 
This leaves no viscous or mixing mechanism to oppose `burrowing' of the tachocline due to meridional cells driven by baroclinic torque $\nabla p \times \nabla \rho$\citep{Mestel1999}. 
This finding suggests that the \citet{Spiegel1992} scenario of burrowing opposed by latitudinal viscous diffusion, and the \citet{Gough1998} suggestion of that PV mixing opposed burrowing \textit{both fail}.
Finally, by process of elimination, the enhanced memory-induced suppression of momentum transport allows the \citet{Gough1998} suggestion that a residual fossil field in the radiation zone is what ultimately limits tachocline burrowing. 



\section{Drift wave Turbulence in a Stochastic Filed}\label{sec:DWT}
This section focuses on the effect of stochastic fields on zonal flow suppression, such as in the case of RMPs at the edge of tokamak. 
Experimental results shows that pre-L-H transition Reynolds stress bursts drop significantly when RMPs are applied to the edge of DIII-D\citep{kriete2020}.
The power threshold for L-H transition increases, as the normalized intensity of radial RMPs ($\delta B_r/B_0$) increases\citep{Leonard_1991,Gohil_2011,Kaye_2011, Ryter_2013, Mordijck_2015,Scannell_2015, in2017,Schmitz_2019}. 
This paper aims to shed light on these two phenomena, and to address the more general question of Reynolds stress decoherence in a stochastic magnetic field. 

\begin{figure}[h!]
\centering
  \includegraphics[scale=0.13]{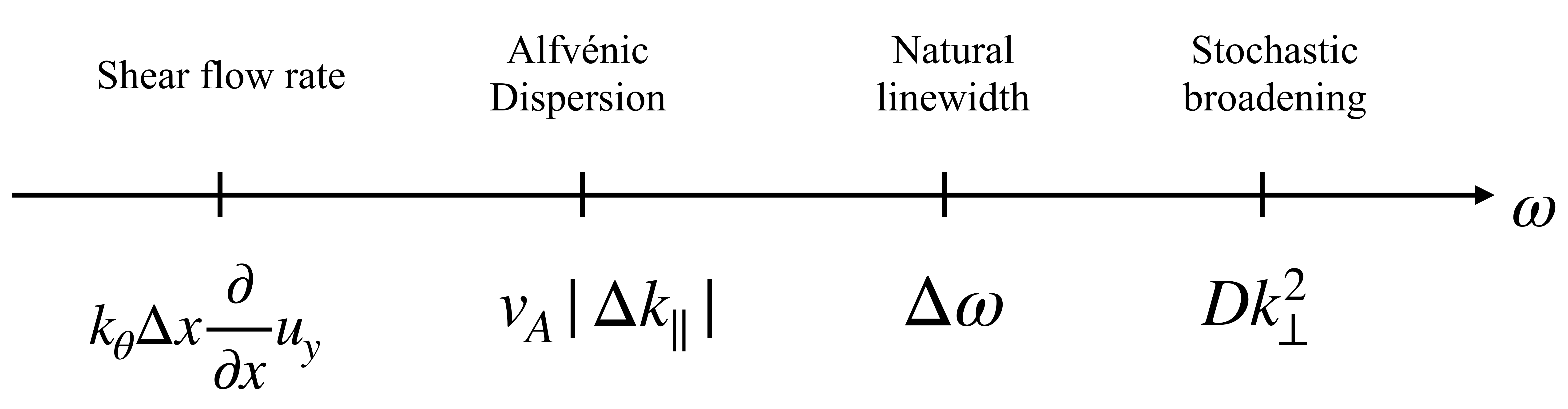}
  \caption{Timescale ordering. 
  We are interested in a regime where stochastic field effect becomes noticeable, which requires $\Delta \omega < D k_\perp^2$. 
 The comparison between Alfv\'enic dispersion rate $v_A |\Delta k_\parallel |$ and stochastic broadening rate $D k_\perp^2$ gives a magnetic Kubo number $Ku_{mag} \simeq 1$. }
  \label{fig:timescale}
\end{figure}

To begin, we explore the timescale ordering for the physics. 
Consider a generalized diffusivity $D_{0}$
\begin{equation}
	D_{0} = Re \{ \sum\limits_{k} \int d\omega \frac{k_\theta^2}{B_0^2} |\phi_{k\omega}|^2 \frac{i}{\omega - v_A k_z + iDk^2} \}
\end{equation}
where the $D$ is a spatial diffusivity under the influence of stochastic field, defined as $D \equiv v_A D_M$, and $v_A \equiv B_0/\sqrt{\mu_0 \rho}$ is Alfv\'en speed\citep{zel1957}.
As discussed below, $v_A$ appears as the characteristic velocity for signal propagation along the stochastic field, since zonal flows follow from the need to maintain $\underline{\nabla} \cdot \underline{J} =0 $, in the face of ambipolarity breaking due to polarization fluxes.
Here $D_M \simeq l_{ac} b^2$ (hearafter $b^2 \equiv \langle B_{st,\perp}^2 \rangle/B_0^2$) is the magnetic diffusivity, first derived by \citet{Rosenbluth_1966}.
Here, the bracket average is a stochastic ensemble average  $\langle \rangle \equiv \int dR^2 \int dB_{st} \cdot P_{(B_{st,x},B_{st,y})} \cdot F$ similar to the bar average in Sec. \ref{sec: Cal. Re. of beta-plane}.
But here $dR^2$ is an averaging area (at scale $1/k_{st}$) over $y$- and $z$- direction. 
$|\phi_{k\omega}|^2$ is the electric potential spectrum, such that 
\begin{equation}
	|\phi|_{k\omega}^2 = \phi_{0}^2  S_{(k)}   \frac{|\Delta \omega_k|}{(\omega - \omega_{0,k})^2 +(\Delta \omega_k)^2},
\end{equation}
where $S_{(k)}$ is the k-spectrum of the potential field, $ \omega_{0,k}$ is the centroid of the frequency spectrum, and $|\Delta \omega|$ is the natural linewidth of potential field.
Performing the frequency integration, we have 
\begin{eqnarray}
	D_{0} &= Re \{ \sum\limits_{k}  \phi_0^2   S_{(k)}   \int d\omega   \{ \frac{i}{(\omega-\omega_{0,k})  + i |\Delta \omega_k|} \frac{i}{\omega - v_A k_z + iDk^2} \} 
	\nonumber
	\\
	&= Re \{ \sum\limits_{k} \phi_0^2 S_{(k)}   \frac{- 2 \pi i}{\omega_{0,k} + i |\Delta \omega_k| -v_A k_z + i D k^2}   \}  .
\end{eqnarray}
Now consider a Lorentzian k-spectrum
\begin{equation}
	S_{(k)} =  \frac{S_0}{(k-k_{0} )^2+ (\Delta k_{\parallel})^2} .
\end{equation}
We have
\begin{eqnarray}
	D_{0} 
	&= Re \{ \int dk_{\parallel}  \phi_0^2  \frac{S_0}{(k-k_{0} )^2+ (\Delta k_{\parallel})^2}
	 \cdot \frac{- 2 \pi i }{\omega_{0,k} -v_A k_z + i |\Delta \omega_k| + i D k^2}   \} 
	\nonumber
	\\
	& \simeq Re \bigg( S_0
	\phi_0^2  (2 \pi)^2  
	\frac{i}{ 
	\omega_{0,k_0}  -k_{0,z} v_A + i|\Delta k_{\parallel}| v_A 
	+ i|\Delta \omega_{k_0}| + i D k_{\perp}^2  	
	} 
	\bigg) ,
	\label{eq: important rate}
\end{eqnarray}
assuming $|\Delta k_{\parallel}| \ll k_{\perp}$ and $\partial \Delta \omega / \partial k \simeq 0 $.
The ordering of these broadenings ($|\Delta k_{\parallel}| v_A $, $|\Delta \omega_{k_0}|$, and $D k_{\perp}^2 $) in the denominator is the key to quantifying stochastic field effects. 
The first term, $|\Delta k_{\parallel}| v_A $, is the bandwidth of an Alfv\'en wave packet excited by drift-Alfv\'en coupling.
Here $v_A |\Delta k_\parallel| \lesssim  v_A/Rq$, where $R$ is major radius and $q\equiv rB_t/RB_p$ is the safety factor.
The bandwidth $|\Delta k_{\parallel}| v_A$ is a measure of the dispersion rate of an Alfv\'en wave packet. 
The second term is the rate of nonlinear coupling or mixing---due to ambient electrostatic micro-instability $|\Delta \omega_{k_0} |\equiv \Delta \omega \simeq \omega_* = k_\theta \rho_s C_s/L_n $, where the $\omega_*$ is drift wave turbulence frequency, $\rho_s $ is gyro-radius, $C_s$ is sound speed, and $L_n$ is density scale length. 
$\Delta \omega$ is comparable to $k_\perp^2 D_{GB}$, where $D_{GB} \equiv \omega_*/k_\perp^2 \simeq \rho_s^2 C_s/L_n$ is the gyro-Bohm diffusivity (for $k_\theta \rho_s \sim 1$).
The third is the stochastic field scattering rate $D k_{\perp}^2  \simeq k_\perp^2 v_A D_M$. 
Ultimately, we will show that $k_{\perp}^2 v_A D_M \gtrsim \Delta\omega_k $ (or $v_A D_M > D_{GB}$) is required for Reynolds stress decoherence (Fig. \ref{fig:timescale}). 
In practice, this occur for $k_{\perp}^2 v_A D_M \gtrsim v_A|\Delta k_\parallel |$, i.e. $Ku_{mag} \simeq1$ is required. 
The condition  $k_{\perp}^2 v_A D_M > \Delta\omega_k $ requires that \textit{stochastic field broadening exceeds the natural turbulence linewidth} \cite{Schmitz_2019}, so that $k_\perp^2 v_A D_M > \Delta \omega$.
Satisfying this requires $b^2  > \sqrt{\beta}  \rho_*^2 \epsilon/q \sim 10^{ -7
}$, where $l_{ac} \simeq R q$, $\epsilon \equiv L_n/R \sim 10^{-2}$, $\beta \simeq 10^{-2\sim -3}$, and normalized gyro-radius $\rho_* \equiv \rho_s/ L_n \simeq 10^{-2 \sim -3}$.
It is believed that $b^2$ at the edge due to RMP is $\sim 10^{-7}$ for typical parameters; hence, the stochastic broadening effect is likely sufficient to dephase the Reynolds stress. 
Following from this condition, we propose a dimensionless parameter $\alpha \equiv b^2 q / \rho_*^2\sqrt{\beta} \epsilon$---defined by the ratio $k_{\perp}^2 v_A D_M /\Delta\omega_k $---to quantify the broadening effect. 
The increment in L-I and I-H power thresholds as $\alpha$ varies are explored using a \textit{modified} Kim-Diamond L-H transition model\citep{kim-diamond2003} in Sec. \ref{sec: kim-diamond}.
We also give a physical insight into stress decoherence by showing how stochastic fields break the `shear-eddy tilting feedback loop', which underpins zonal flow growth by modulational instability. 

\subsection{Model Setup}
\begin{figure}[h!]
\centering
  \includegraphics[scale=0.16]{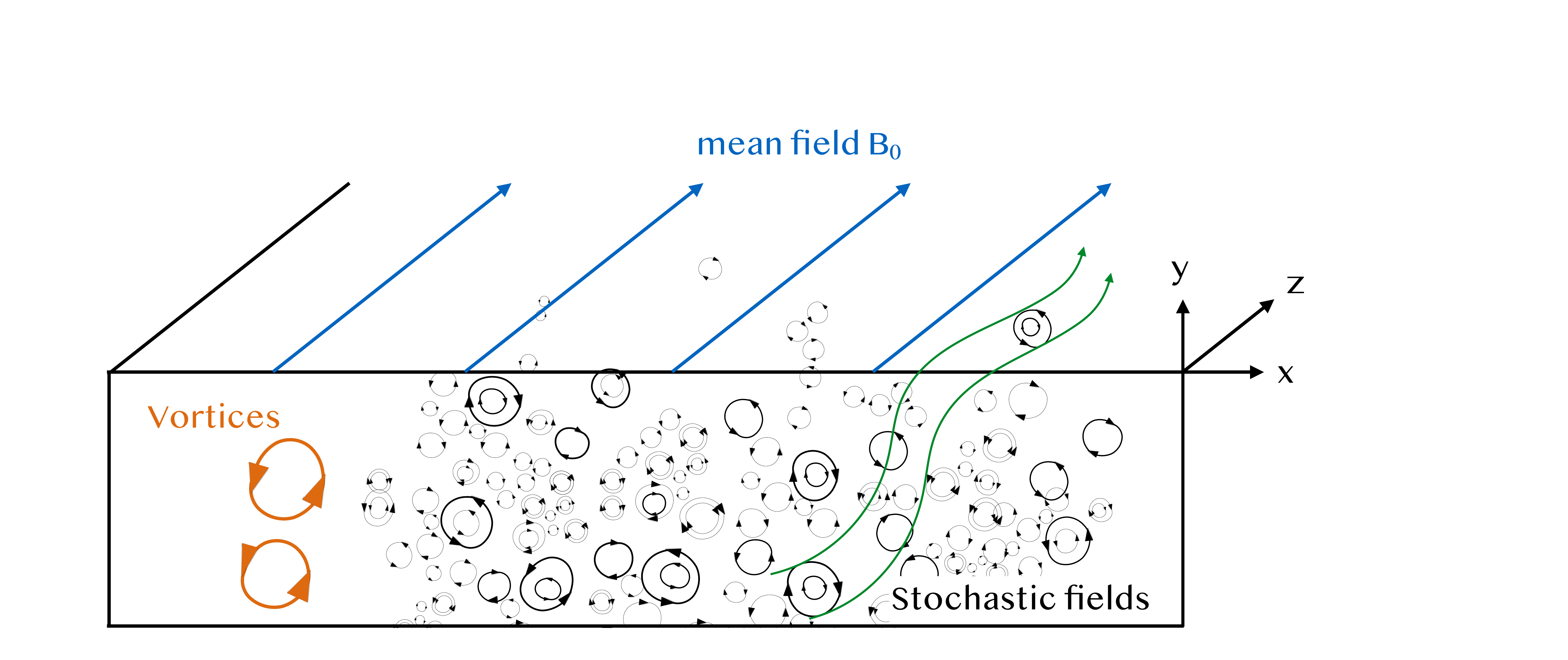}
  \caption{Magnetic fields at the edge of tokamak. RMP-induced stochastic fields (black loops) lie in radial ($x$) and poloidal ($y$) plane.  Mean toroidal field is treading through stochastic fields perpendicular in $z$-direction (blue arrows).}
  \label{fig:3D}
\end{figure}

We construct a model in Cartesian (slab) coordinates---$x$ is radial, $y$ is poloidal, and $z$ is toroidal direction, in which the mean toroidal field lies (Fig. \ref{fig:3D}). 
A current flows in the toroidal direction, producing a mean poloidal field.
In contrast to the tachocline, here the magnetic field is 3D, and stochasticity results from the overlap of magnetic islands located at the resonant $\underline{k}\cdot \underline{B} =0$ surfaces. The stochasticity is attributed to the external RMP field, and typically occurs in a layer around the separatrix. 
The distance between neighboring magnetic field trajectories diverges exponentially, as for a positive Lyapunov exponent. 
Stochastic fields due to RMPs resemble Zel'dovich `cells'\citep{Zel1983_mag_percolation} (Fig. \ref{fig:channel}), lying in $x-y$ plane with a mean toroidal field (on $z$-axis), threading through perpendicularly.
Of course, once overlap occurs, the coherent character of the perturbations is lost, due to finite Kolmogorov-Sinai entropy (i.e. there exists a positive Lyapunov exponent for the field). 
In this case, the magnetic Kubo number is modest $Ku_{mag} \lesssim 1$.

We start with 4 field equations---

1. Vorticity evolution---$\underline{\nabla} \cdot \underline{J} =0$
\begin{equation}
	\frac{\partial}{\partial t} \zeta_z + u_y \frac{\partial}{\partial y} \zeta_z + u_z \frac{\partial}{\partial z} \zeta_z 
	= \frac{1}{\rho} B_0 \frac{\partial}{\partial z} J_z + \frac{1}{\rho} B_{x,st} \frac{\partial}{\partial x} J_z + \frac{2 \kappa}{\rho}\frac{\partial}{\partial y} P,
	\label{eq1}
\end{equation}
where $\zeta$ is the vorticity, $u_y$ is $E \times B $ shear flow, $u_z$ is intrinsic rotation, and $\kappa$ is curvature. 

2. Induction evolution
\begin{equation}
	\frac{\partial}{\partial t} A_z + u_y \frac{\partial}{\partial y} A_z 
	= -\frac{B_{x,st}}{B_0} \frac{\partial}{\partial x} \phi - \frac{\partial}{\partial z} \phi + \eta \nabla^2 A_z,
	\label{eq2}
\end{equation}
where $\phi $ is electric potential field ($\zeta \equiv \nabla \times v = \frac{1}{B_0} \nabla^2 \phi$). 

3. Pressure evolution
\begin{equation}
 \frac{\partial}{\partial t}p  + (\bold{u} \cdot \nabla) p = - \gamma p (\nabla \cdot \bold{u}),
 \label{eq3}
\end{equation}
where $\gamma$ is the adiabatic index .

4. Parallel acceleration 
\begin{equation}
	\frac{\partial}{\partial t} u_z + (\bold{u} \cdot \nabla) u_z = - \frac{1}{\rho} \frac{\partial}{\partial z} p,
	\label{eq4}
\end{equation}
where $p$ is pressure. 
\subsection{Calculation and Results}\label{sec: kim-diamond}
We define a Els\"asser-like variable $f_{\pm, k\omega} \equiv \widetilde{\phi}_{k\omega} \pm v_A   \widetilde{A}_{k\omega}$, and combine Eq. (\ref{eq1}) and (\ref{eq2}) to obtain
\begin{equation}
\begin{split}
	 &(-i \omega + \langle u_y\rangle  ik_y) f_{\pm, k\omega} \pm v_A (ik_z + ik_j \frac{B_{j,st}}{B_0}  ) f_{\pm, k\omega}
	 \\ 
	&= \frac{\widetilde{u}_x}{k^2} \frac{\partial}{\partial x} \nabla^2 \langle \phi \rangle + \frac{2 \kappa}{\rho}  ik_y (\frac{B_0}{-k^2}) \widetilde{p}  \equiv S_f,
	\label{eq:9}
\end{split}
\end{equation}
where $S_f$ is the source function for $f_{\pm, k\omega}$.
Eq. \ref{eq:9} is the evolution equation for the Els\"asser response to a vorticity perturbation. 
Note that this response is defined by 

1. Propagation along the total magnetic field, i.e. $ik_z + ik_j B_{j,st}/B_0$. Note this includes propagation along the wandering magnetic field component. 

2. Advection by mean flow $ik_y \langle u_y \rangle$.

3. Finite frequency $i \omega$.

Hence, the Els\"asser response for $f_{\pm, k\omega}$ is be obtained by integrating along trajectories of total magnetic field lines (including perturbations), i.e. 
\begin{equation}
	 f_{\pm, k\omega} = \int d\tau e^{i(\omega- \langle u_y \rangle k_y \mp v_A k_z ) \tau}  e^{\mp i v_A \int d \tau^{\prime} (\frac{B_{i,st}}{B_0} k_i)} \times S_f
\end{equation}
Integration along the perturbed field trajectory can be implemented using the stochastic average over an scale ($1/ k_{st}$)
\begin{equation}
	 e^{\mp i v_A \int d \tau^{\prime} (\frac{B_{i,st}}{B_0} k_i)} \rightarrow \langle e^{\mp i v_A \int d \tau^{\prime} (\frac{B_{i,st}}{B_0} k_i)} \rangle ,
	\label{eq:11}
\end{equation}
where the bracket denotes an average over random radial excursions $\delta x_i = \int d\tau v_A B_{i,st}/B_0$.
This yields the Els\"asser response
\begin{equation}
	\langle f_{\pm, k\omega} \rangle = \int d\tau e^{i(\omega- \langle u_y \rangle k_y \mp v_A k_z ) \tau} \langle e^{\mp i v_A \int d \tau^{\prime} (\frac{B_{i,st}}{B_0} k_i)} \rangle  \times S_f,
	\label{eq:11}
\end{equation}
where $i$, $j $ are indexes for perpendicular components and $Dk^2 = D_x k_x^2 + D_y k_y^2$ and $M_f$ is the propagator.
Here, $\langle e^{\mp i v_A \int d \tau^{\prime} (\frac{B_{i,st}}{B_0} k_i)} \rangle$ is set by the diffusivity tensor $\underline{\underline{D}} = v_A^2  \int d\tau^"  b^2_{j,st (\tau^")}  $ so 
\begin{equation}
           \langle e^{\mp i v_A \int d \tau^{\prime} (\frac{B_{i,st}}{B_0} k_i)} \rangle
	    \simeq 1 - k_i D_{ij} k_j \tau
	    \simeq  e^{-\underline{k} \cdot \underline{\underline{D}} \cdot \underline{k} \tau} ,
	   \label{eq:13} 
\end{equation}
where $\tau$ is the decorrelation time due to field stochasticity, such that $\tau \simeq \int d\tau^" \simeq  l_{ac}/ v_A$.
We assume no correlation between $x$- and $y$-direction of stochastic field (i.e. and $\langle B_{x,st } B_{y,st } \rangle =0 $) and $\langle B_{i,st} \rangle=0$.
Hence, only diagonal terms of $\underline{\underline{D}}$ survive (i.e. $D_{ij} = \delta_{ij }v_A l_{ac} b_i^2$).
A number of important comments are in order here. 
First, $D\simeq v_A D_M$, indicating that vorticity response decorrelation  occurs by Alfv\'enic pulse diffusion along wandering magnetic fields. 
This is a consequence of the fact that PV (or polarization charge) perturbations (which determine the PV or polarization charge flux---i.e. the Reynolds force) are determined via $\underline{\nabla}\cdot\underline{J} = 0$, the characteristic signal speed for which is $v_A$.
Second, $v_A D_M$ is actually \textit{independent of $B_0$} and is a set only by $b^2$.
To see this, observe that $b^2 \equiv \langle B_{st}^2 \rangle/B_0^2$,  $v_A = B_0/\sqrt{\mu_0 \rho }$, and $l_{ac} = Rq$.
Thus, $D \propto b^2$ reflects the physics that decorrelation occurs due to pulses traveling along stochastic fields, \textit{only}. 
In this respect, the result here closely resembles the 2D case (i.e. $\beta$-plane MHD) discussed in Section \ref{sec1}.
Third, $v_A$ for the mean field enters only via the linear vorticity response---which is used to compute the vorticity flux---and thus the Reynolds force. 

Now we have the averaged Els\"asser response
\begin{equation}
	\langle f_{\pm, k\omega} \rangle
	= \frac{i}{(\omega- \langle u_y \rangle k_y \mp v_A k_z) + iDk^2} \times S_f
	\equiv M_f S_f,
\end{equation}
where $Dk^2 = D_x k_x^2 + D_y k_y^2$.
And $M_f$ is a propagator defined as 
\begin{equation}
	M_f = \frac{1}{2} \bigg(
	\frac{i}{( \omega_{sh} - v_A k_z) + iDk^2} + \frac{i}{( \omega_{sh} + v_A k_z) + iDk^2}
	\bigg),
\end{equation}
where $ \omega_{sh} \equiv\omega - \langle u_y \rangle  k_y $ is the shear flow Doppler shifted frequency.
From Eq. (\ref{eq:9}), we have the fluctuating vorticity
\begin{equation}
	\widetilde{\zeta} = \frac{1}{B_0}\nabla^2 \widetilde{\phi}= \sum\limits_{k\omega} Re [   M_f (\frac{-k^2}{B_0} S_f) ]
\end{equation}
Hence, the response of vorticity ($\widetilde{\zeta}$) to the vorticity gradient and curvature term in the presence of stochastic fields is:
\begin{eqnarray}
	\widetilde{\zeta} =   \sum\limits_{k\omega} \Big[
	Re(M_f) (-\frac{\widetilde{u}_{x,k\omega} }{B_0} \frac{\partial}{\partial x} \nabla^2 \langle \phi \rangle)
	+ Re\big( ik_y M_f \frac{2 \kappa}{\rho}   \widetilde{p}_{k\omega} \big)
	\Big] 	
	\label{eq:19}
\end{eqnarray}
The first term determines the diffusive flux of vorticity.
The second sets the off-diagonal stress, or \textit{residual stress}, that depends on the pressure perturbation and the curvature of the mean magnetic field.
We calculate the residual stress term in Eq. (\ref{eq:19}) by using another set of Els\"asser-like variables $g_{\pm, k\omega} \equiv \frac{\widetilde{p}_{k\omega}}{\rho C_s^2} \pm \frac{\widetilde{u}_{z,k\omega}}{C_s}$, defined from Eq. (\ref{eq3}) and (\ref{eq4}), and follow the approach discussed above. 
This yields $\widetilde{p}_{k\omega} = M_g \widetilde{u}_x \frac{\partial}{\partial x} \langle p \rangle$ and $M_g$ is defined as
\begin{equation}
	M_g  
	= \frac{1}{2} \bigg(
	\frac{i}{( \omega_{sh} - C_s k_z) + i D_s k^2} + \frac{i}{( \omega_{sh} + C_s k_z) + i D_s k^2}
	\bigg)
	\simeq \frac{i}{\omega_{sh}}. 
\end{equation}
where $D_{s}  \equiv C_s D_M$ (for pressure decorrelation rate $\tau_c = l_{ac}/C_s$) is the diffusivity due to an acoustic signal propagating along stochastic fields. 
Notice that $\widetilde{p}$ is the pressure perturbation set by the acoustic coupling.
Hence, it has slower speed $C_s \ll v_A$ (or $\beta \ll 1$) as compared to Alf\'enic coupling. 
An ensemble average of total vorticity flux yields
\begin{equation}
\begin{split}
	\langle \widetilde{u}_x \widetilde{\zeta} \rangle =   
	&-
	\sum\limits_{k \omega} |\widetilde{u}_{x,k\omega} |^2 Re(M_f) \frac{\partial}{\partial x}  \langle \zeta \rangle
	\\
	&+  \underbrace{
	\sum\limits_{k \omega}  \bigg[ |\widetilde{u}_{x,k\omega} |^2 Re(i k_y M_f M_g)  \frac{2 \kappa}{\rho}  \frac{\partial}{\partial x } \langle p \rangle 
	\bigg]
	}_{\textit{Residual Stress}}.
\end{split}
\end{equation}
Notice that $D_s k^2 \simeq C_s D_M k^2$.
Hence, the broadening effect of random acoustic wave propagation itself is negligible as compared to the natural linewidth, since the plasma $\beta \ll 1$.
Now, we have 
\begin{equation}
	\langle \widetilde{u}_x \widetilde{\zeta} \rangle =   
	-
	D_{PV}
	 \frac{\partial}{\partial x}  \langle \zeta \rangle
	+
	F_{res}
	\kappa \frac{\partial}{\partial x } \langle p \rangle ,
\end{equation}
where $D_{PV}\equiv \sum\limits_{k \omega} |\widetilde{u}_{x,k\omega} |^2 Re(M_f)$ is PV diffusivity, and  $F_{res} \simeq \sum\limits_{k \omega} \frac{-2k_y}{\omega_{sh} \rho} D_{PV,k\omega}$ is the residual stress.
Notice that there is no parity issue lurking in the term $2 k_y/ \omega_{sh} \rho$ since $2 k_y/\omega_{sh}\rho \propto 2 \cancel{k_y}/\cancel{k_y}\rho \propto 2/\rho$ (i.e. even) for $k_y \langle u_y \rangle \ll \omega \simeq \omega_* $.
By using the Taylor Identity\citep{taylor1915}, we rewrite the PV flux as Reynolds force $\langle \widetilde{u}_x \widetilde{\zeta} \rangle = \frac{\partial}{\partial x} \langle  \widetilde{u}_x \widetilde{u}_y\rangle $.
In the limit of the $D_{PV}$ and $F_{res}$ slowly varying as compared with vorticity $\langle \zeta \rangle$ and pressure $\langle p \rangle$, respectively, the poloidal Reynolds stress is
\begin{equation}
	\langle \widetilde{u}_x \widetilde{u}_y \rangle =   
	- D_{PV}  \frac{\partial}{\partial x} \langle u_y \rangle 
	+ F_{res} \kappa  \langle p \rangle,
	\label{eq: Reynolds stress DF}
\end{equation}
where the effective viscosity is 
\begin{equation}
	D_{PV} = \sum\limits_{k \omega} |\widetilde{u}_{x,k\omega} |^2 \frac{v_A b^2  l_{ac} k^2}{\omega_{sh}^2 + (v_A  b^2  l_{ac}  k^2)^2} .
	\label{eq:D_pv}
\end{equation}
This indicates that \textit{both the PV diffusivity and residual stress (and thus the Reynolds stress) are suppressed as the stochastic field intensity $b^2$ increases}, so that $v_A b^2 l_{ac}  k^2$ exceeds $\omega_{sh}$. 
This result is consistent with our expectations based upon scaling and with the Reynolds stress burst suppression in presence of RMPs, observed in \citet{kriete2020}. 
This model is build on gyro-Bohm scaling and hence the stochastic dephasing effect is insensitive to the details of the turbulence mode (e.g. ITG, TEM,…etc.), within that broad class. 

\begin{figure}[h!]
\centering
  \includegraphics[scale=0.15]{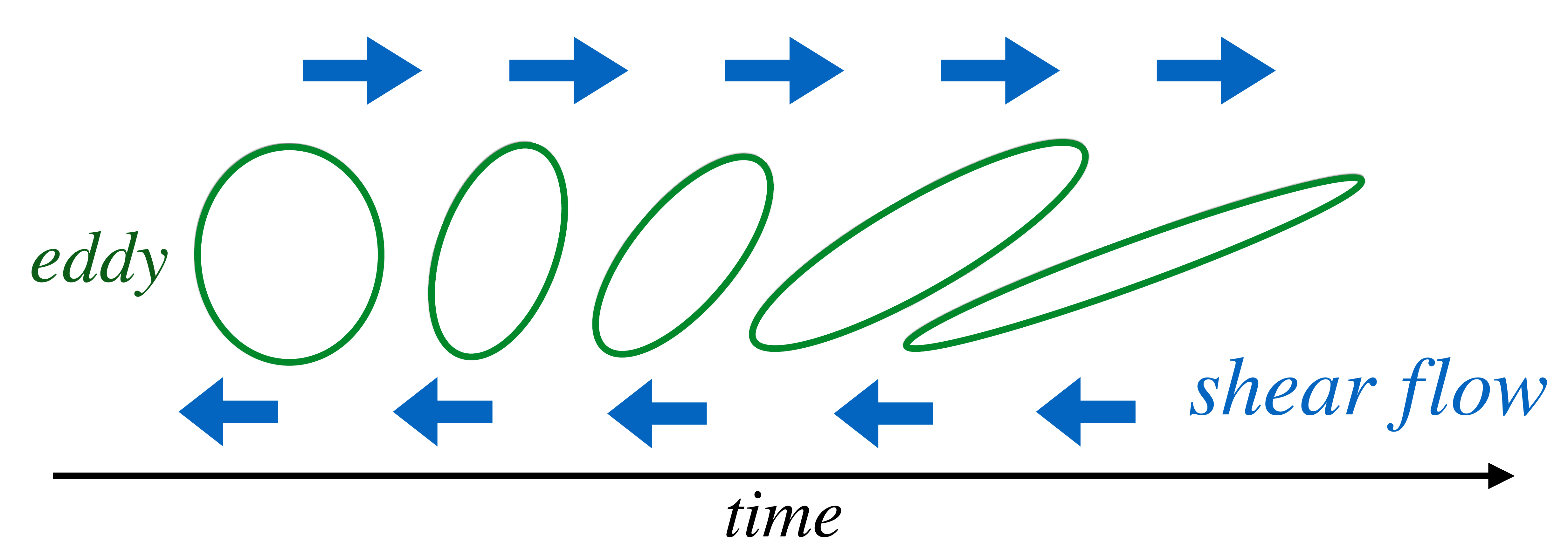}
  \caption{Shear-eddy tilting feedback loop. The $E\times B$ shear generates the $\langle k_x k_y \rangle$ correlation and hence support the non-zero Reynolds stress. And the Reynold stress, in turns, modifies the shear via momentum transport. Hence, the shear flow reinforce the self-tilting. }
  \label{fig:eddy_tilting}
\end{figure}

Physical insight into the physics of Reynolds stress decoherence can be obtained by considering the effect of a stochastic magnetic field on the `shear-eddy tilting feedback loop'. 
Recall that the Reynolds stress is given by 
\begin{equation}
	\langle \widetilde{u}_x \widetilde{u}_y \rangle 
	=  - \sum\limits_k \frac{| \widetilde{\phi}_k |^2}{B_0^2} \langle k_y k_x \rangle.
\end{equation}
Thus, a non-zero stress requires $\langle  k_y k_x \rangle \neq 0$, i.e. a spectrally averaged wave vector component correlation. 
This in turn requires a spectral asymmetry. 
In the presence of a seed shear, $k_x$ tends to align with $k_y$, producing $\langle \rangle \neq 0$ (Fig. \ref{fig:eddy_tilting}).
To see this, observe that Snell's law states
\begin{equation}
\frac{ d k_x}{dt} =-\frac{\partial (\omega_{0,k} + k_y u_y)}{\partial x}
\simeq 0 -  \frac{\partial ( k_y  u_y)}{\partial x}.
\end{equation}
So, to set a non-zero phase correlation $\langle  k_y k_x \rangle \neq 0$, we take $k_x \simeq k_x^{(0)} - k_y \frac{\partial  \langle u_y\rangle}{\partial x} \tau_c$, where $\tau_c$ is a ray scattering time that limits ray trajectory time integration.
Ignoring $k_x^{(0)}$, we then find 
\begin{equation}
	\langle \widetilde{u}_x \widetilde{u}_y \rangle 
	\simeq 0  + \sum\limits_k  \frac{| \widetilde{\phi}_k |^2}{B_0^2} k_y^2 \frac{\partial \langle u_y\rangle}{\partial x} \tau_{c,k}.
\end{equation}
Note that the existence of correlation is unambiguous, and the Reynolds stress is manifestly non-zero.
Here, \textit{eddy tilting} (i.e. $k_x$ evolution) has aligned wave vector components.
Once $\langle u_x u_y \rangle \neq 0$, flow evolution occurs due to momentum transport. 
Then, flow shear amplification further amplifies the Reynolds stress, etc. 
This process constitutes the `shear-eddy tilting feedback loop', and underpins modulational instability amplification of zonal shears.
Central to shear-eddy tilting feedback is the proportionality of stress cross-phase to shear. 
However, in the presence of stochastic fields, the correlation $\langle k_x k_y \rangle $ is altered. 
To see this, consider drift-Alf\'en turbulence, for which
\begin{equation}
	\omega^2 - \omega_* \omega -k_{\parallel}^2 v_A^2 = 0.
	\label{eq: drift-alfven}
\end{equation}
Let $\omega_0 $ be the frequency of the drift wave roots. 
Now, let $k_\parallel = k_\parallel^{(0)} + \underline{k}_\perp \cdot(\underline{B}_{st,\perp}/B_0)$ due to stochastic field wandering, and $\delta \omega$ the corresponding ensemble averaged correction to $\omega_0$---i.e. $\omega = \omega_0 + \delta \omega$.
After taking an ensemble average of random fields from Eq. (\ref{eq: drift-alfven}), we obtain $ \langle \delta \omega \rangle \simeq v_A^2  (2 k_{\parallel} \frac{\langle \underline{B}_{st,\perp} \rangle }{B_0}  \cdot \underline{k}_{\perp} + \langle(\frac{\underline{B}_{st,\perp}}{B_0}  \cdot \underline{k}_{\perp})^2 ) \rangle/\omega_0 
$, where $\langle B_{i,st} \rangle=0 $ so the first term vanishes.
The ensemble averaged frequency shift is then 
\begin{equation}
	\langle \delta \omega \rangle
	\simeq \frac{1}{2} \frac{v_A^2}{\omega_0}  b^2 k_{\perp}^2.
\end{equation}
Here, $\langle \omega_0\rangle \simeq \omega_*$, corresponding to the drift wave. 
Note that $\delta \omega \propto \langle B_{st}^2 \rangle $ is independent of $B_0$, except for $\omega_0$. 
Thus, in the presence of shear flow, the Reynolds stress becomes
\begin{equation}
	\langle \widetilde{u}_x \widetilde{u}_y \rangle \simeq    \sum\limits_k  \frac{| \widetilde{\phi}_k |^2}{B_0^2} 
(k_y^2 \frac{\partial  \langle u_y\rangle}{\partial x} \tau_{c,k}
+\frac{1}{2} k_y \frac{v_A^2 k_{\perp}^2 }{\omega_0} \frac{\partial b^2}{\partial x} \tau_{c,k}).
\end{equation}
This indicates that for $\frac{\partial  \langle u_y\rangle}{\partial x} < \frac{v_A^2 k_{\perp}^2 }{\omega_0} \frac{\partial b^2}{\partial x} $, the shear-eddy tilting feedback loop is broken, since the $\langle k_x k_y \rangle$ correlation is no longer set by flow shear. 
In practice, this requires $b^2 \gtrsim 10^{-7}$, as deduced above. 


We modify a well-known predator-prey model of the L-H transition, the Kim-Diamond model\citep{kim-diamond2003} to include the effects of stochastic fields. 
The Kim-Diamond model is a zero-dimensional reduced model, which evolves fluctuation energy, Reynolds stress-driven flow shear, and the mean pressure gradient.
As heat flux is increased, a transition from L-mode to Intermediate phase (I-phase) and to H-mode occurs. 
Here, we include the principal stochastic field effect---Reynolds stress decoherence. 
This is quantified by the dimensionless parameter $\alpha \equiv  q b^2/\sqrt{\beta}\rho_*^2 \epsilon  $ derived in Sec. \ref{sec:DWT}. 
The aim is to explore the changes in L-H transition evolution (i.e. power threshold increment) due to magnetic stochasticity. 
This dimensionless parameter $\alpha$ quantifies the strength of stochastic dephasing relative to turbulent decorrelation.
As shown in the previous paragraph, the $E \times B$ shear feedback loop that forms the zonal flow is broken by the stochastic fields. 
Hence, the modification enters the \textit{shear decorrelation term} in the turbulence ($\xi$) evolution, 
the corresponding term in the zonal flow energy ($v_{ZF}^2 $) evolution, 
and the pressure gradient ($\altmathcal{N}$) evolution. 
The third term is smaller by $\sqrt{\beta}$ (i.e. $\alpha \rightarrow \alpha\sqrt{\beta}$), due to the fact that \textit{acoustic} wave scatting is what causes decoherence in the pressure evolution. 
A factor $1/(1+ c\alpha)$ captures the modification due to the effect of stochastic suppression effect, where $c$ a constant. 
The \textit{modified} Kim-Diamond model becomes
\begin{eqnarray}
	&\frac{\partial \xi}{\partial t} = \xi \altmathcal{N} - a_1 \xi^2 
	- a_2 (\frac{\partial \langle u_y\rangle}{\partial x})^2 \xi 
	-\underbrace{a_3 v_{ZF}^2 \xi \cdot \frac{1}{(1+ a_4\alpha)}
	}_{\textit{Reynolds stress decoherence}}
	\\
	&\frac{\partial v_{ZF}^2}{\partial t} 
	= \underbrace{a_3 v_{ZF}^2 \xi \cdot \frac{1}{(1+ a_4\alpha)}
	}_{\textit{Reynolds stress decoherence}}
	-b_1 v_{ZF}^2
	\\
	&\frac{\partial \altmathcal{N}}{\partial t} 
	= -\underbrace{c_1 \xi \altmathcal{N}\cdot \frac{1}{(1+ a_4 \alpha \sqrt{\beta})} 
	}_{\textit{turbulent diffusion of pressure}}
	-c_2 \altmathcal{N}+Q,
\end{eqnarray}
where $a_i $, $b_i$, and $c_i$ ($a_1 =0.2$, $a_2 =0.7$, $a_3 =0.7$, $a_4 =1$, $b_1 =1.5$, $c_1 =1$, $c_2 =0.5$, $\sqrt{\beta}$=0.05) are model-dependent coefficients, and $Q$ is the input power. 

\begin{figure}[h!]
\centering
  \includegraphics[scale=0.25]{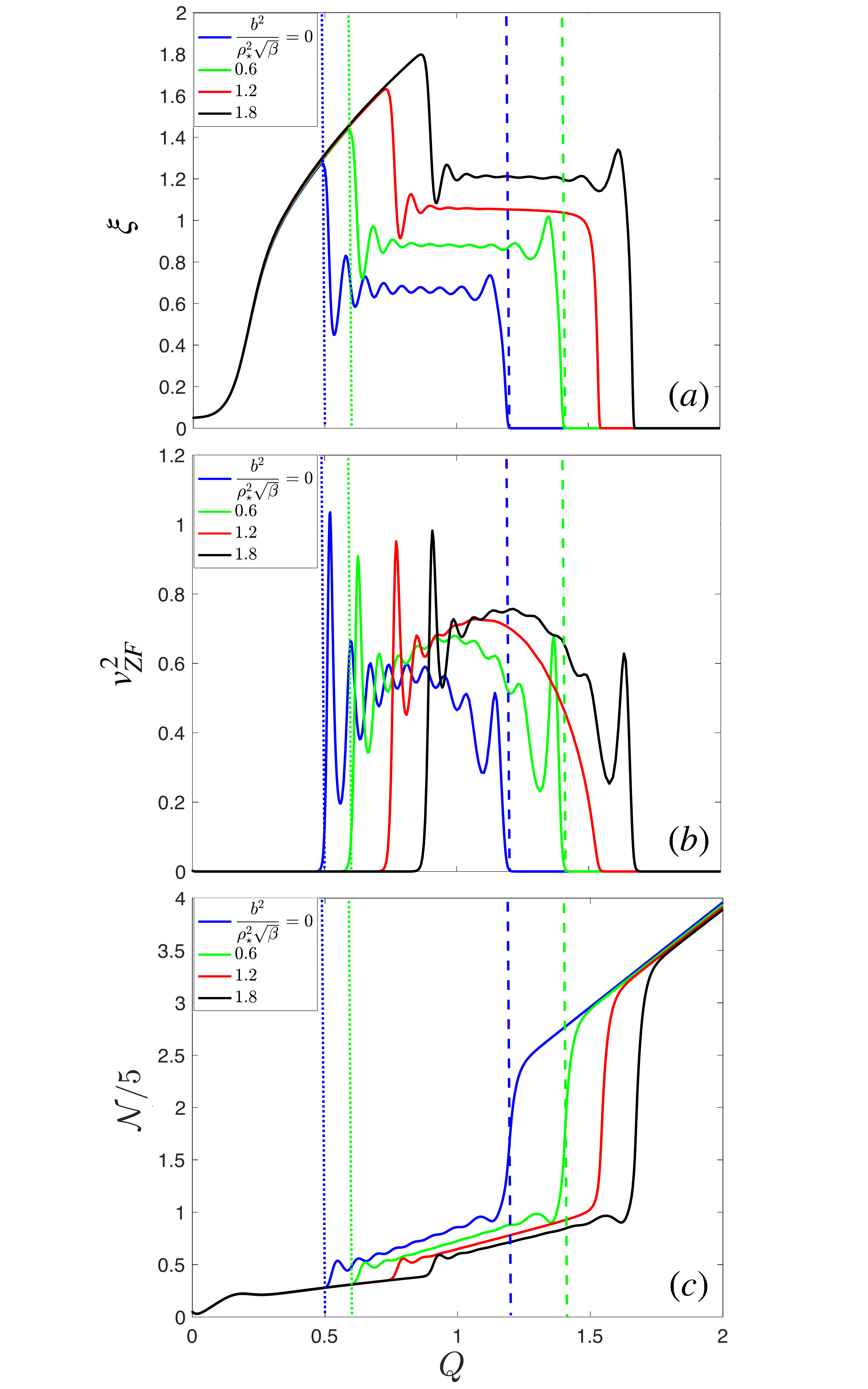}
  \caption{Modified Kim-Diamond model. (a) Turbulent intensity $\xi$. (b) Zonal flow energy $v_{ZF}^2$. (c) Pressure gradient $\altmathcal{N}$ evolution with increasing input power $Q$.
  Dotted lines indicate L-I transitions, dashed lines indicate I-H transitions.  
  As we increase the mean-square stochastic field ($b^2$), L-I and I-H transitions power threshold shift to the right, i.e. from $ b^2 / \rho_*^2 \sqrt{\beta} = 0$ (blue) to $0.6$ (green). }
  \label{fig:3v}
\end{figure}

\begin{figure}[h!]
\centering
  \includegraphics[scale=0.22]{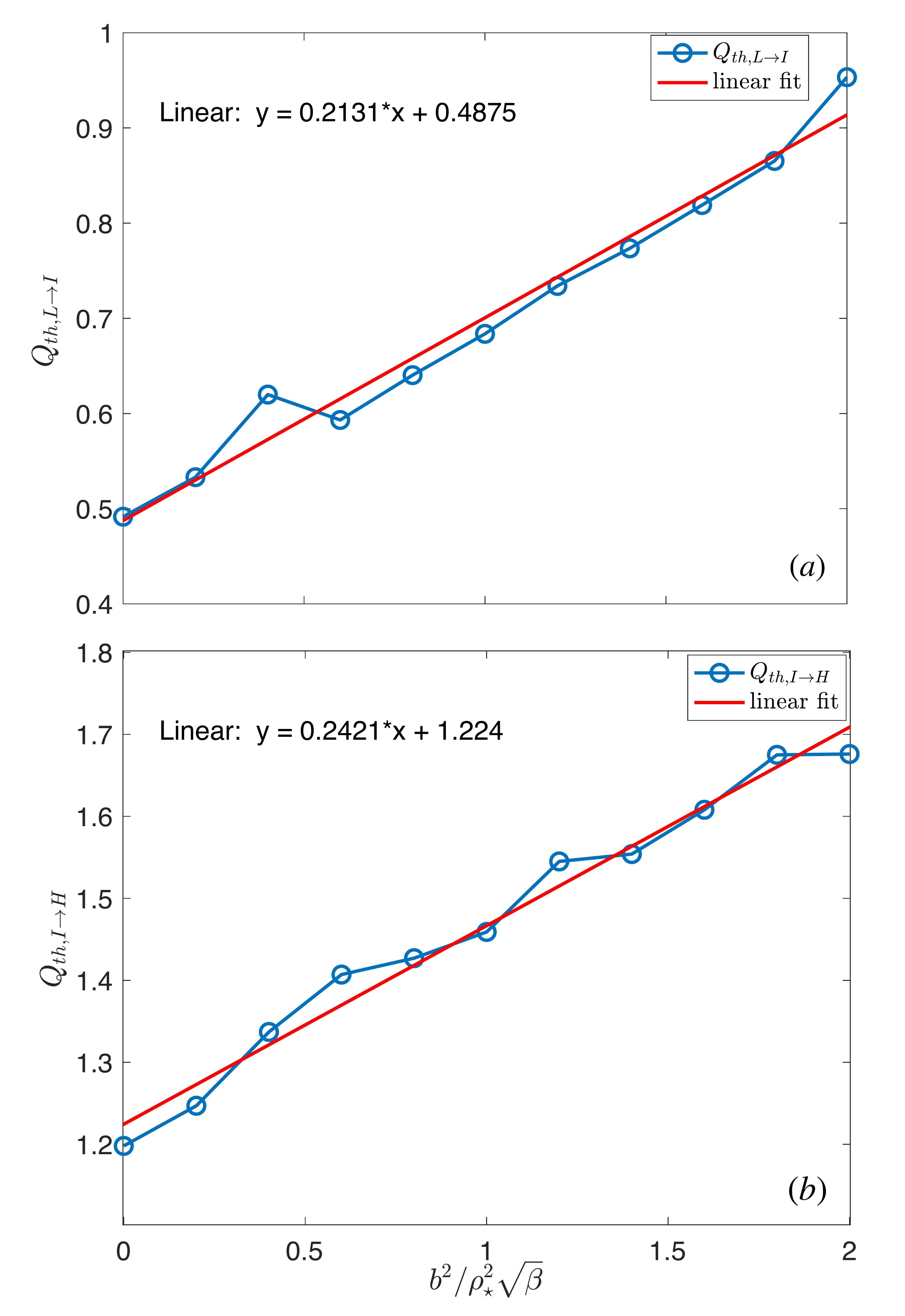}
  \caption{Power threshold increments in modified Kim-Diamond model. (a) L-I transition power threshold increment. (b) I-H transition power threshold increment.
     Mean-square stochastic fields ($b^2$) shift L-H and I-H transition thresholds to higher power, in proportional to $ b^2/\rho_*^2 \sqrt{\beta}$.}
  \label{fig:LH_increment}
\end{figure}

We find that stochastic fields raise the L-I  and I-H transition power thresholds, linearly in proportion to $\alpha$ (Fig. \ref{fig:LH_increment}). 
And recall that $\alpha$ is proportional to stochastic fields intensity $b^2$ (Fig. \ref{fig:3v}).
This is a likely candidate to explain the L-H power threshold increment in DIII-D \citep{Schmitz_2019}.

We are also interested in stochastic field effects on the toroidal Reynolds stress $\langle \widetilde{u_x} \widetilde{u_z}\rangle$, which determines intrinsic toroidal rotation.
Consider toroidal Eq. (\ref{eq4}) with the stochastic fields effect $\frac{\partial}{\partial z} = \frac{\partial}{\partial z}^{(0)} + \underline{ b} \cdot \nabla_{\perp}$. 
We have
\begin{equation}
	\frac{\partial}{\partial t} \langle u_z \rangle  +  \frac{\partial}{\partial x} \langle \widetilde{u}_x \widetilde{u}_z \rangle  = - \frac{1}{\rho} \frac{\partial}{\partial x} \langle b \widetilde{p} \rangle,
	\label{eq: toroidal equation}
\end{equation}
The second term on the LHS is the toroidal Reynold stress $\langle \widetilde{u_r} \widetilde{u_z}\rangle$. 
The RHS contains the $\langle b\widetilde{p} \rangle$ the kinetic stress.
Both of these terms can be dephased by stochastic fields, but the dephasing of the former is of primary importance.  
In the context of intrinsic rotation, we follow the method for the derivation of decoherence of the poloidal residual stress---i.e. using El\"asser-like variables $g_{\pm, k\omega} \equiv \frac{\widetilde{p}_{k\omega}}{\rho C_s^2} \pm \frac{\widetilde{u}_{z,k\omega}}{C_s}$ from Eq. (\ref{eq3}) and (\ref{eq4}).
The only difference from the previous residual stress calculation is the presence term of
$ \frac{\partial}{\partial x} \langle u_z\rangle$, 
and hence the source of toroidal stress becomes $S_{g,\pm} \equiv -\frac{\widetilde{u}_{x,k\omega}}{\rho C_s^2} \frac{\partial}{\partial x} \langle p \rangle \mp \frac{\widetilde{u_x}}{C_s} \frac{\partial}{\partial x} \langle u_z \rangle$.
We find $\widetilde{u}_{z,k\omega} = M_{g,-}C_s$, where $M_{g,-}$ is a propagator such that 
\begin{equation}
	M_{g,-}  = \frac{i}{2} \big(
	\frac{S_{g,+}}{(\omega_{sh} - C_s k_z) + i D_s k^2} - \frac{S_{g,-}}{(\omega_{sh} + C_s k_z) + i D_sk^2}
	\big),
\end{equation}
Noted that when $ \frac{\partial}{\partial x} \langle u_z\rangle =0$, the propagator $M_{g,-}$ reduces to $M_{g}$. 
Thus, the toroidal Reynold stress is 
\begin{equation}
\begin{split}
	\langle \widetilde{u}_x \widetilde{u_z} \rangle 
	&=   
	\sum\limits_{k\omega} |\widetilde{u}_{x,k\omega}|^2  \big[
	\frac{
	 - 2D_s  k^2  }
	{\omega_{sh} ^2 + (2   D_s  k^2  )^2} \frac{\partial  \langle u_z\rangle}{\partial x}
	\\
	&+ 
	\frac{
	-   2D_s  k^2 }
	{ \omega_{sh} ^2 + (2  D_s  k^2  )^2} \frac{ k_z}{\omega_{sh}\rho }  \frac{\partial  \langle p\rangle}{\partial x}
\big].
\end{split}
\label{eq: z-stress}
\end{equation}
The first term in RHS contains the \textit{turbulent viscosity} ($\nu_{turb}$), which we define as
\begin{equation}
\begin{split}
	\nu_{turb} 
	&\equiv \sum\limits_{k\omega} |\widetilde{u}_{x,k\omega}|^2  \frac{
	  2D_s  k^2  }
	{ \omega_{sh} ^2 + (2   D_s  k^2  )^2} 
	\\
	&= \sum\limits_{k\omega} |\widetilde{u}_{x,k\omega}|^2  \frac{
	  2C_s b^2 l_{ac} k^2  }
	{ \omega_{sh} ^2 + (2  C_s b^2 l_{ac}   k^2  )^2}.
\end{split}
	\label{eq: z-turbulent viscosity}
\end{equation}
This turbulent viscosity has a form similar to $D_{PV}$ in Eq. (\ref{eq:D_pv}). 
However, decorrelation of $\nu_{turb}$ is set by $C_s$ while that of $D_{PV}$ is set by $v_A$. 
Thus, decoherence effects here are weaker. 
The second term in Eq. (\ref{eq: z-stress}) contains the \textit{toroidal residual stress} ($F_{z,res}$)
\begin{equation}
	F_{z,res} \equiv \sum\limits_{k\omega} (\frac{- k_z}{\omega_{sh} \rho})  |\widetilde{u}_{x,k\omega}|^2  \frac{
	(  2D_s  k^2) }
	{ \omega_{sh}^2 + (2  D_s  k^2  )^2}  
	\sim \sum\limits_{k\omega} \frac{- k_z}{\omega_{sh} \rho} \nu_{turb, k\omega}.
	\label{eq: z-residual stress}
\end{equation}
Notice that non-zero value of $F_{z,res}$ requires symmetry breaking (i.e. $\langle k_z k_y\rangle \neq 0$) since
$\frac{k_z }{ \omega_{sh} \rho}  \propto \frac{k_z }{ k_y } $.
Thus, \textit{a symmetry breaking condition---non-zero  $ \langle k_z k_y \rangle$---must be met for finite residual toroidal residual stress ($F_{z,res}$) }.
Here, $\langle k_z k_y \rangle$ must now be calculated in the presence of the stochastic field.
The details of this calculation involve determining the interplay of stochastic field effects with spectral shifts (i.e. symmetry breaking by $E \times B $ shear) and inhomogeneities (i.e. spectral symmetry breaking by intensity gradient).
This will involve competition between the radial scale length of stochastic fields and the scales characteristic of the spectral shift (induced by $E \times B $ shear) and the spectral intensity gradient. 
This detailed technical study is left for a future publication. 
We rewrite the toroidal stress as
\begin{equation}
	\langle \widetilde{u}_x \widetilde{u_z} \rangle 
	=   - \nu_{turb} \frac{\partial}{\partial x} \langle u_z\rangle + F_{z,res} \frac{\partial}{\partial x} \langle p\rangle,
\end{equation}
which has similar form to that of poloidal Reynolds stress in Eq. (\ref{eq: Reynolds stress DF}).
This shows that stochastic fields reduce the toroidal stress and hence slow down the intrinsic rotation. 
However, from Eq. (\ref{eq: z-turbulent viscosity}) and (\ref{eq: z-residual stress}), the stochastic suppression effect on toroidal stress and residual stress depends on $C_s D_M$ (not $v_A D_M$), and so is weaker than for zonal flows. 

\section{Discussion} \label{sec: discussion}
In general terms, we see that 42 years after the influential paper by \citet{Rosenbluth1977}, the physics of plasma dynamics in a stochastic magnetic field remains theoretically challenging and vital to both cosmic and magnetic fusion energy (MFE) plasma physics. 
Transport in a state of coexisting turbulence and stochastic magnetic field is a topic of intense interest. 
In this paper, we discussed aspects of momentum transport and zonal flow generation in two systems with low effective Rossby number, where dynamics evolve in the presence of a stochastic magnetic field. 

The first system is the solar tachocline--- with weak mean magnetization, strong magnetic perturbation, and $\beta$-plane MHD dynamics. 
Here, a tangled magnetic network generated by fluid stretching at large $Rm$ defines an effective resisto-elastic medium in which PV transport occurs.
We show that coupling to bulk elastic waves, with frequency $\omega^2 \simeq \overline{B_{st}^2}k^2/\mu_0 \rho$, results in decoherence of the PV flux and Reynolds force, thus limiting momentum transport.
Moreover, this effect sets in for seed field energies well below that required for Alfv\'enization.
Physically, the stress decoherence occurs via coupling of fluid energy to the elastic network of fields, where it is radiatively dissipated. 
One implication of this prediction of quenched momentum transport is that tachocline burrowing cannot be balanced by momentum transport. 
This bolsters the case for Gough and McIntyre's suggestion\citep{Gough1998} that a fossil magnetic field in radiation zone is what ultimately limits meridional cell burrowing. 

The second system is the L-mode tokamak edge plasma, in the presence of a stochastic magnetic field induced by external RMP coils. 
Here, the system is 3D, and field lines wander due to islands overlap. 
The magnetic Kubo number is modest. 
We showed that the `shear-eddy tilting feedback loop' is broken by a critical $b^2$ intensity, and that $ k_\perp^2 v_A D_M$ characterizes the rate of stress decoherence. 
Note that the Alfv\'en speed follows from charge balance, which determines Reynolds stress.
A natural threshold condition for Reynolds stress decoherence emerges as $k_\perp^2 v_A D_M/\Delta \omega >1$. 
In turn, we show that this defines a dimensionless ratio $\alpha$, which quantifies the effect on zonal flow excitation, and thus power thresholds.
$\alpha \simeq 1$ occurs for $b^2 \simeq 10^{-7}$, consistent with stochastic magnetic field intensities for which a significant increment in power threshold occurs. 
Note that this scaling is somewhat pessimistic (i.e. $\rho_*^{-2}$).

This study has identified several topics for future work. 
These include developing a magnetic stress---energy tensor evolution equation, for representing small-scale fields in real space. 
Fractal network models of small-scale magnetic field are promising in the context of intermittency.
A better understanding of stochastic field effects on transport for $Ku_{mag} \geq1$ is necessary as a complement to our $Ku_{mag} \leq 1$ model-based understanding. 
For MFE plasmas, an 1D model for the L-H transition evolution is required. 
This study will introduce a new length scale (M. Jiang \& W. Guo et al. in press), which quantifies the radial extent of the stochastic region.
Finally, the bursty character\citep{kriete2020} of pre-transition Reynolds work, suggests that a statistical approach to the transition is required. 
The challenge here is to identify the physics of the noise and flow bursts, and how the presence of stochasticity quenches them. 
The stochasticity-induced change in `shear-eddy tilting feedback loop' discussed herein is a likely candidate for the quenching of the noise and flow burst. 
\begin{acknowledgments}
We thank Lothar Schmitz, D. M. Kriete,  G. R. McKee, Zhibin Guo, Gyungjin Choi, Weixin Guo, and Min Jiang for helpful discussions.
Simulating discussions at the 2019 Aix Festival de Th\'eorie are also acknowledged. 
This research was supported by the US Department of Energy, Office of Science, Office of Fusion Energy Sciences, under award No. DE-FG02-04ER54738.
\end{acknowledgments}
\section*{Data Availability}
The data that support the findings of this study are available from the corresponding author upon reasonable request.


\section*{References}
\bibliography{aipsamp}

\end{document}